\documentclass[fleqn,12pt,twoside]{article}
\usepackage{espcrc1}

\usepackage{graphicx}
\usepackage[figuresright]{rotating}

\newcommand{\AmS}{{\protect\the\textfont2
  A\kern-.1667em\lower.5ex\hbox{M}\kern-.125emS}}

\title{Exotic nuclear phases in the inner crust of neutron stars in the
       light of Skyrme-Hartree-Fock theory}

\author{P. Magierski\address{Faculty of Physics, Warsaw University of Technology,
        \\
        ul. Koszykowa 75, 00-662 Warsaw, Poland}%
        \thanks{This research was supported in part by the Polish Committee
        for Scientific Research (KBN) under Contract No.~5~P03B~014~21
        and the Wallonie/Brussels-Poland
        integrated action program. Numerical calculations were performed
        at the Interdisciplinary Centre for Mathematical and Computational
        Modelling (ICM) at Warsaw University.},
        A. Bulgac\address{Department of Physics,  University of Washington, \\
                          Seattle, WA 98195--1560, USA}
	 and
        P.-H. Heenen\address{Service de Physique Nucl\'{e}aire Th\'{e}orique, \\
	                     U.L.B - C.P. 229, B 1050 Brussels, Belgium.}
			     }

\begin{document}

\maketitle

\begin{abstract}
The bottom part of the neutron star crust is
investigated using the Skyrme-Hartree-Fock approach
with the Coulomb interaction treated beyond the Wigner-Seitz
approximation.
A variety of nuclear phases is found to coexist
in this region. Their stability and relative energies are
governed by the Coulomb, surface and shell energies. We have
also found that a substantial contribution is coming from the
spin-orbit interaction.
\end{abstract}

\bigskip

The crust of a neutron star with a typical mass of $1.4~{\rm M}_\odot$,
contains only $1.2\%$ of the total stellar mass.
Nevertheless its structure is important for
understanding of several observational issues and
thus provides a motivation for theoretical studies
(see e.g. \cite{bbp,nvb,lpr,wko,lfb,oya1,pra,dhm,dha,mh,mbh}
and references therein).

According to various theoretical models
in the inner part of the crust, due to high density and pressure,
the nuclei forming a crystalline lattice are immersed in a neutron gas.
The electrons at these densities are ultrarelativistic particles
and are assumed to form a uniform background.

In the bottom part of the inner crust
the interplay between the Coulomb and the surface energies
leads to the appearance of various nuclear phases,
characterized by different shapes.
The early predictions, based on the liquid drop or Thomas-Fermi models,
with quite restrictive symmetry conditions,
lead to the considerations of five phases, formed
in the region where the nucleon density varies from $0.03$ to $0.1$ fm$^{-3}$.
These are: spherical nuclei, rods (``spaghetti'' phase), slabs (``lasagna'' phase),
tubes and bubbles (see \cite{pra} and references therein).

These approaches however provided an oversimplified description
of the system. First, due to the Wigner-Seitz approximation which
has been used, there is no distinction between phases resulting from
different lattice geometries, second, the quantum
effect associated with the shell correction energy has been neglected,
and last but not least, the symmetry conditions, which admit only spherical,
cylindrical and planar structures, are not sufficient to describe properly
the variety of nuclear shapes present in the crust.

Our approach, which is free from the aforementioned limitations, can shed
some light on the complexity of the inner crust structure.
The ground and excited (isomeric) states of neutron-proton-electron ($npe$) matter
have been determined by the Skyrme-Hartree-Fock method. We have solved
the HF equations by discretization in coordinate space,
within a cubic box and with  periodic
boundary conditions imposed on the nucleon wave functions \cite{bfh}.
The details of this approach have been described in Ref.\cite{mh}.

In the HF approximation, the total energy of a nuclear system is, in general,
a functional of three local densities:
total nucleon density $\rho ({\bf r})=\rho_{p}({\bf r})+\rho_{n}({\bf r})$,
total kinetic density $\tau ({\bf r})=\tau_{p} ({\bf r})+\tau_{n} ({\bf r})$ and
total spin-orbit density ${\bf J}({\bf r})={\bf J}_{p}({\bf r})+{\bf J}_{n}({\bf r})$.
Each of the densities is in turn expressed through the single-particle wave
functions $\phi_{i}^{q}( {\bf r}, \sigma )$:
\begin{equation}
\rho_{q}({\bf r})=
\sum_{i,\sigma}n_{i}^{q}\mid\phi_{i}^{q}( {\bf r}, \sigma )\mid^{2},
\end{equation}
\begin{equation}
\tau_{q} ({\bf r})=
\sum_{i,\sigma}n_{i}^{q}\mid\nabla\phi_{i}^{q}( {\bf r}, \sigma )\mid^{2},
\end{equation}
\begin{equation}
{\bf J}_{q}({\bf r})=
-i\sum_{i,\sigma , \sigma '}n_{i}^{q}
\phi_{i}^{q*}( {\bf r}, \sigma )\nabla\phi_{i}^{q}( {\bf r}, \sigma ')\times
\langle\sigma\mid\mbox{\boldmath $\sigma$}\mid\sigma '\rangle ~~~~(q=\mbox{n,p}),
\end{equation}
where $ \frac{1}{2}\sigma$ ($\sigma =\pm 1$) is the spin projection
on the $z$-axis, and $n_{i}^{q}$ is the occupation factor, which
is $1$ below the Fermi level and zero otherwise.

The Skyrme energy functional can be written as:
\begin{eqnarray}
E_{sk}&=&
\int \Bigg{[}
\displaystyle{\frac{\hbar^{2}}{2m} }
\tau+B_{1}\rho^{2}+B_{2}(\rho_{p}^{2}+\rho_{n}^{2})+
B_{3}\rho\tau+B_{4}(\rho_{p}\tau_{p}+\rho_{n}\tau_{n})+ \nonumber \\
&+& B_{5}\rho\nabla^{2}\rho+B_{6}(\rho_{p}\nabla^{2}\rho_{p}+\rho_{n}\nabla^{2}\rho_{n} )+
B_{7}\rho^{2+\gamma}+ \nonumber \\
&+&B_{8}\rho^{\gamma}(\rho_{p}^{2}+\rho_{n}^{2})+B_{9}(\rho\nabla\cdot{\bf J}+
\rho_{p}\nabla\cdot{\bf J}_{p}+\rho_{n}\nabla\cdot{\bf J}_{n})
\Bigg{]} d^{3}r.
\end{eqnarray}
The coefficients $B_{i}$ depend on the Skyrme
force parameters \cite{bfh}. It is useful to distinguish the following
contributions to the Skyrme energy functional:
\begin{eqnarray}
E_{sksurf}&=&\int \left (
B_{5}\rho\nabla^{2}\rho+B_{6}(\rho_{p}\nabla^{2}\rho_{p}+\rho_{n}\nabla^{2}\rho_{n} )
\right )
d^{3}r , \\
E_{so}&=&\int \left (
B_{9}(\rho\nabla\cdot{\bf J}+
\rho_{p}\nabla\cdot{\bf J}_{p}+\rho_{n}\nabla\cdot{\bf J}_{n}) \right )
d^{3}r ,
\end{eqnarray}
where $E_{sksurf}$ contains the contribution to the surface energy
and $E_{so}$ is the spin-orbit term.

For a correct description of $npe$ matter one has to add to the Skyrme energy functional
the Coulomb term and the kinetic energy of electrons:
\begin{equation} \label{ehf}
E = E_{sk} + E_{Coul} + E_{el},
\end{equation}
where the direct Coulomb energy has been calculated
by solving the Poisson equation
with periodic boundary conditions. The Coulomb exchange term
for both protons and electrons has been
treated in the Slater approximation.

Typical results for two values of
the average total density
$\rho_{av}$ $=$ $\displaystyle{\frac{1}{V}\int_{V} (\rho_{p} + \rho_{n})} d^{3}r$,
and the proton-to-nucleon ratio ($Z/A$) ensuring the
$\beta$ stability condition,
are shown in Fig. 1. The total energy densities
have been plotted with respect to the value of the
spherical system $Q_{p}=0$.

\begin{figure}[ht]
\includegraphics[angle=-90,scale=0.6]{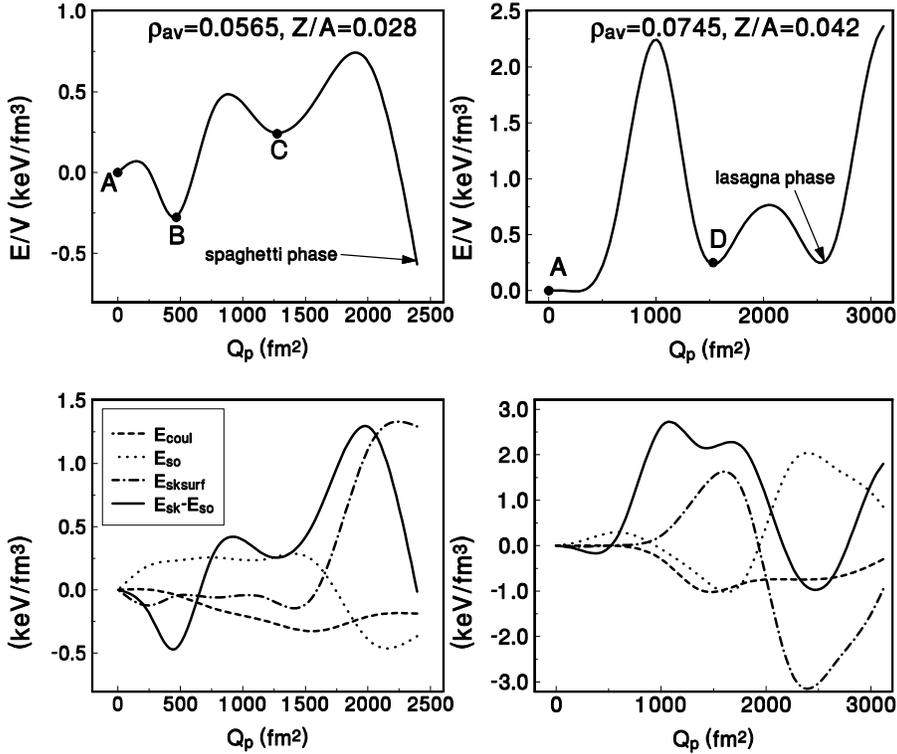}
\caption{The results of the Skyrme-Hartree-Fock calculations
(using SLy4 force) presented
as a function of the axial proton quadrupole moment: $Q_{20}=Q_{p}$.
The calculations have been performed in a box of lengths: $d=26 fm$
(left panel) and $d=20.8 fm$ (right panel).
The upper subfigures show the total energy density and the lower subfigures
the contribution to the total energy coming from various terms in the
density functional (see text for details).
The phase A consists of spherical nuclei in the scc lattice,
the phases B and C denote the scc geometries differing by
the nuclear deformations. The D phase corresponds to the
bcc lattice.}
\end{figure}

The energy density has several
local minima, corresponding to various nuclear shape
and lattice geometries.
It was shown in Ref.\cite{mh} that the relative position of various
nuclear configurations
is a sensitive function of the density. It is a consequence of
the shell effects associated with unbound neutrons \cite{bma1}.
Hence different nuclear phases may coexist
in the same density range. There is an obvious larger variety
of nuclear shapes than it was
predicted in the previous approaches (\cite{pra,dha} and
references therein).

In the lower subfigures of Fig. 1, the contributions
from the Coulomb energy, the $E_{sksurf}$ term
and the spin-orbit term, have been shown.
One can make the following observations:
\begin{itemize}
\item The Coulomb energy decreases as a function of $Q_{p}$
for small values of quadrupole moment and then increases slightly
for large $Q_{p}$ values. This is related to the fact that large
$Q_{p}$ values correspond to the configuration where the nuclei
from neighboring cells join each other. Hence the configuration
favored by the Coulomb interaction consists of deformed nuclei,
but not necessarily in the shape of rods or slabs.
\item The part of the surface energy coming from the $E_{sksurf}$
term follow approximately the behavior of the term $E_{sk}-E_{so}$.
It suggests that the surface energy dominates
the behavior of $E_{sk}$ as a function of the deformation.
\item The amplitude of variation of the spin-orbit term $E_{so}$
 is of the same order as the $E_{sk}-E_{so}$.
 The spin-orbit energy is a sensitive
 function of $Q_{p}$ and clearly have a direct influence on the appearance
 and depths of local minima of the total energy curve.
 One has to remember that this term gives the substantial contribution
 to the shell energy.
\end{itemize}

Finally one has to mention that the pairing correlation were
not included in the above calculations. The pairing would
smooth the total energy curve and thus its role in the inner
edge of the neutron star crust needs a thorough examination.

\end{document}